\def\sign{\mathop{\rm sign}\nolimits}
\def\Tr{\mathop{\rm Tr}\nolimits} 
\def\bra#1{\langle #1 \,\vert}
\def\ket#1{\vert\, #1 \rangle}
\def\frac#1#2{{#1 \over #2}}
\documentstyle{espart}
\begin{document}
\begin{frontmatter}
\title{Electromagnetic Nucleon Properties and Quark Sea
Polarization}
\author[Bochum]{Chr.V.~Christov\thanksref{Sofia}\thanksref{EmChr}}
\author[Krakow]{A.Z.~G\'orski\thanksref{EmGor}}
\author[Bochum]{K.~Goeke\thanksref{EmGoe}} and
\author[StPetrs]{P.V.~Pobylitsa}
\address[Bochum]{Institut f\"ur Theoretische Physik II,
Ruhr-Universit\"at Bochum, D-44780 Bochum, Germany}
\address[Krakow]{Institute of Nuclear Physics, Radzikowskiego 152, 31-342
Cracow, Poland}
\address[StPetrs]{Petersburg Nuclear Physics Institute, Gatchina,
St.Petersburg 188350, Russia}
\thanks[Sofia]{Permanent address:Institute for Nuclear Research and Nuclear
Energy, Sofia, Bulgaria}
\thanks[EmChr]{christov@neutron.tp2.ruhr-uni-bochum.de}
\thanks[EmGor]{gorski@bron.ifj.edu.pl}
\thanks[EmGoe]{goeke@hadron.tp2.ruhr-uni-bochum.de}

\begin{abstract}
In this paper we present the derivation as well
as the numerical results for all electromagnetic form factors
of the nucleon within the chiral quark soliton model in the  semiclassical
quantization scheme. The model is based on a semibosonized
Nambu -- Jona-Lasinio lagrangean where the boson fields are treated as
classical ones. Other observables, namely the nucleon
mean squared radii, the magnetic moments and the nucleon--$\Delta$ splitting
are calculated as well. The calculations have been done taking into
account the quark sea polarization effects.
The final results, including rotational $1/N_c$ corrections, are
compared with the existing experimental data and they are found to be
in a good agreement for the constituent quark mass of $400-420$ MeV.
\end{abstract}
\end{frontmatter}
\vskip-16cm
\hskip10.3cm{RUB-TPII-2/94}
\vskip-0.2cm
\hskip10.3cm{May, 94}
\newpage

\section{Introduction}

In the last years one of the most challenging problems in elementary
particle physics seems to be the solution of QCD in the low energy region.
The main difficulties are due to the non-perturbative effects caused by the
growing effective coupling constant of the fundamental theory
in the low energy limit. This prevents one from using the well-known
main tool of theoretical physics --- the perturbation theory.
Because of this, the most intriguing features of QCD -- confinement and
chiral symmetry breaking -- still remain conceptual and practical
problems.

The above mentioned obstacles have initiated an increasing interest
among the physicists in non-perturbative methods and effective
low-energy models of hadrons. The effective models are expected to
mimic the behaviour of QCD at energies below $\sim 1~GeV$ (confinement
and/or chiral symmetry breaking) and to reproduce experimental data in
this region. In principle, these models could be related to QCD by
integrating out gluonic fields and reparameterizing fermionic
degrees of freedom.

The simplest purely fermionic Lorentz invariant model with spontaneous
chiral symmetry breaking is the Nambu--Jona-Lasinio (NJL)
model ~\cite{Nambu61}. It contains chirally invariant local four-fermion
interaction terms. The NJL Lagrangian in its simplest SU(2) form has the
following structure:
\begin{equation}
{\cal L}_{NJL} = \bar \Psi \, (i \FMSlash\partial - m_0)\, \Psi +
{G\over 2} \, [(\bar \Psi \Psi )^2 + (\bar \Psi i \vec\tau
\gamma_{5} \Psi )^2\,] \ ,
\label{NJLLAG} \end{equation}
where $G$ is the coupling constant, $m_0$ is the current
quark mass and $\vec\tau$ are the Pauli matrices in the isospin
space. We consider the $up$ and $down$ quarks to be degenerated
in mass.

The NJL model is generally solved after applying the well known
bosonization procedure following Eguchi~\cite{Eguchi74} to arrive at
the model expressed in terms of auxiliary meson fields
$\sigma, {\vec\pi}$:
\begin{equation}
{\cal L}_{NJL} = \bar \Psi \  [ \ i \FMSlash\partial - m_0-
g_\pi (\sigma + i\, {\vec\pi\cdot\vec\tau}\, \gamma_5 )\ ] \, \Psi -
{\mu^2\over 2} (\sigma^2 + {\vec\pi}^2).
\label{NEWLAG} \end{equation}
The new coupling constant $\mu^2$ is related to
the initial $G$ by $G=g_\pi^2/\mu^2$. Here, $g_\pi$ is the physical
pion--quark coupling constant implying that $\vec\pi$
is the physical pion field.
The meson fields are constrained to the chiral circle:
\begin{equation}
\sigma^2({\vec x}) + {\vec\pi}^2({\vec x}) = f^2_{\pi} \ ,
\label{CHIRCIR} \end{equation}
where $f_{\pi}=93$ MeV is the pion decay constant.

In the chiral quark soliton model~\cite{Diakonov89,Reinhardt88,TMeissner89}
based on the lagrangean (\ref{NEWLAG}) (frequently referred simply as NJL
model) the baryons appear as a bound state of $N_c$ valence quarks
coupled to the polarized Dirac sea. Operationally the
baryon sector of the model is solved in two steps. In the first step,
motivated by the large $N_c$ limit, a static localized solution
(soliton) is found  by solving the corresponding equations of motion
in an iterative self-consistent~\cite{TMeissner89}  procedure assuming that
the $\sigma$ and $\vec\pi$ fields have hedgehog structure. However, this
hedgehog soliton does not preserve the spin and isospin. In order to describe
the nucleon properties one  needs nucleon states with the proper spin
and isospin numbers. To this end making use of the rotational zero modes
the soliton is quantized. In fact, such a
cranking procedure was elaborated in refs.~\cite{Adkins83,Diakonov89} and
problems arising in this scheme due to the regularization are considered at
first in ref.~\cite{Reinhardt89}.  Within the semiclassical
quantization scheme quite successful calculations including Dirac sea
polarization effects have been reported for the nucleon-delta mass
splitting~\cite{Goeke91}, magnetic moments~\cite{Wakamatsu91} and
axial-vector form factor~\cite{TMeissner91} as well as some results for the
nucleon electric form factors~\cite{Gorski92}. Very recently, in the
semiclassical quantization procedure important $1/N_c$ rotational
corrections have been derived~\cite{Christov94} which improve
considerably the theoretical values for the isovector magnetic moment.

Because of several reasons the model, described by lagrangean
(\ref{NEWLAG}), is considered as one of the most promising effective
theories describing low energy QCD phenomena.

First, the model is the simplest quark model which provides mechanism
for spontaneous breaking of the chiral symmetry --- the basic
feature of QCD. The mesons appear as composite states of quark-antiquark
pairs. The philosophy behind this approach is
based on the hypothesis that the chiral symmetry breaking is the dominant
mechanism which determines the structure of the low-mass hadrons whereas
the confinement has no direct impact om them~\cite{Diakonov89}.

Second, the self-consistent solitonic solution
has been found to exist in the large $N_c$ limit of the model within
the physically acceptable range of parameters~\cite{TMeissner89} fixed
in the meson sector. The chiral soliton provides a good description of
the nucleon and gives us possibility to take into account vacuum
polarization effects from the Dirac quark sea.
Calculations done in the
semiclassical quantization procedure yield quite reasonable results for
quantities like the nucleon radii, $\Sigma$--terms,
axial vector coupling constant, $\Delta$--nucleon splitting as well as the
mass splittings within the octet and decuplet.

Third, there are various hints~\cite{Diakonov84,Ball89,Schaden90}  that the
NJL-type Lagrangian can be obtained from QCD in various low-energy
approximations. It should be stressed that the large $N_c$ limit plays a
prominent role in those considerations.

The aim of this paper is to calculate the electromagnetic nucleon form
factors within the $SU(2)$ chiral quark soliton model based on the
semibosonized NJL-type lagrangean with the vacuum polarization effects taken
into account. The calculations done in the semiclassical quantization scheme
will include the rotational $1/N_c$ corrections,
which have been shown~\cite{Christov94} to be important for the
isovector magnetic moment. Calculations of this sort
give us possibility to calculate the nucleon electromagnetic
form factors, as well as such quantities like the
electric mean squared radii, the magnetic moments, the
nucleon--$\Delta$ energy splitting, and the electric and magnetic
charge distributions of the nucleon.

The paper is organized as follows: we begin in Section 2 with
the electromagnetic current in the model defined in terms of path integrals
and introduce the rotational zero modes treating the angular velocity
as a perturbation. After that we consider the problem of regularization.
In Section 3 we derive the expressions for the form factors including terms
up to the linear order in angular velocity ($1/N_c$).  In Section 4 we
present and discuss our numerical results.

\section{The current and regularized action}

Our main goal in this section is to introduce the matrix element of the
electromagnetic currents in terms of path integrals and to evaluate it
within the semiclassical quantization scheme. We follow the
valence picture for the nucleon in which it appears as a bound state
of $N_c$ valence quarks coupled to the Dirac sea. Since the NJL model is not
renormalizable a special attention will be paid to the problem of
regularization.

We start with the definitions of the
electromagnetic current of a fermion field $\Psi(x)$:
\begin{equation}
J_\mu(x)=\Psi^\dagger(x)\gamma_0\gamma_\mu\hat Q \Psi(x)\ .
\label{CURRENT}\end{equation}
Here $\hat Q$ is the quark charge matrix
\begin{equation}
\hat Q \equiv {1\over 6} \hat1 + {1\over2} \tau^3\ .
\label{CHARGE}\end{equation}
Using the partition function of the model in Minkowski space
\begin{equation}
Z_{NJL} = \int{\cal D}U\,\int{\cal D}\Psi {\cal D}\Psi^\dagger\,
\e^{ \,i \int \d^4x {\cal L}_{NJL} (x) } ,
\label{NEWGEN} \end{equation}
we express the nucleon matrix
element of current $J_\mu$ as a path integral:
\begin{eqnarray}
&&\bra{N({\vec p\prime})}\,J_\mu (0)\,\ket{N({\vec p})}
=\lim\limits_{T\to-i\infty} \,{\frac 1Z} \, \int \d^3x \, \d^3y \
\e^{i{\vec p^\prime\cdot\vec x}} \, \e^{-i{\vec p\cdot\vec y}}
\int{\cal D}U \int {\cal D}\Psi{\cal D}\Psi^\dagger \nonumber \\
&&{}_\times J_N(T/2,{\vec x})J_N^\dagger(-T/2,{\vec y})\,
\Psi^\dagger(0)\gamma_0\gamma_\mu\hat Q\Psi(0)\,\e^{\,
i\int \d^4z\,\Psi^\dagger D(U)\Psi}\,.
\label{Eq2} \end{eqnarray}
written in terms of quark $\, \Psi\ $,$\ \Psi^\dagger\, $ and meson $U$
fields. The equality (\ref{Eq2}) should be understood as a limit at large
Euclidean time separation. Here $Z$ is the normalization factor which
is related to the same path integral but without the electromagnetic
current $J_\mu$. In fact, the latter represents~\cite{Diakonov89} the
correlation function of two nucleon currents $J_N(T/2,{\vec x})$.
$D$ is the operator
\begin{equation}
D(U)=i\partial_t-h(U),
\label{DIRAC}\end{equation}
which includes the one-particle hamiltonian
\begin{equation}
h(U)=\frac {\vec{\alpha}\cdot\vec{\nabla}}i+\beta M U^{\gamma_5}+m_0\beta\,.
\label{HAMIL}\end{equation}
Here $\vec{\alpha}$ and $\beta$ are the Dirac matrices and the constituent
quark mass $M~=~gf_{\pi}$. The meson fields can be equivalently written as
\begin{equation}
U=\ \e^{ i\vec{\tau}\cdot\hat{\vec{\pi}}P(\vec
r)}\,
\label{DEFU}\end{equation}
where
\begin{equation}
\hat{\vec{\pi}}\,=\,\frac {\vec{\pi}}{|\vec{\pi}|}\quad {\rm and} \quad
|\vec{\pi}|\,=\,f_\pi\,{\sin(P(\vec  r))}
\label{PI}\end{equation}
It can be easily checked that the hamiltonian is hermitian: $ h^\dagger =
h$. The composite $N_c$ quark operator with the quantum numbers
$JJ_3,TT_3$ (spin, isospin) of the nucleon can be chosen to be one of the
Ioffe currents~\cite{Joffe81}:
\begin{equation}
J_N(x)=\frac
1{N_c!}\varepsilon^{\beta_1\cdots\beta_{N_c}}\Gamma^{\{f_1\cdots
f_{N_c}\}}_{JJ_3,TT_3}
\Psi_{\beta_1f_1}(x)\cdots\Psi_{\beta_{N_c}f_{N_c}}(x),
\label{Eq6} \end{equation}
where $\beta_i$ is the color index, and $\Gamma^{f_1\cdots f_{N_c}}_{
JJ_3,TT_3}$ is a matrix in $f_i$ standing for both the flavor and
the spin indices.

In eq.(\ref{Eq2}) the quarks can be integrated out:
\begin{eqnarray}
&&\langle N({\vec p^\prime})|\,J_\mu (0)\,\ket{N({\vec p})}\,=\,
{\frac 1Z}\,N_c\,\Gamma^{\{f\}}_{JJ_3,TT_3}\,\Gamma^{\{g\}*}_{JJ_3,TT_3}\,
\,\int \d^3x\,\d^3y\, \e^{i{\vec p^\prime\cdot\vec x}}
\e^{-i{\vec p\cdot\vec y}}\int{\cal D}U\nonumber\\
&&{}_\times\Biggl\{\bra{
T/2,{\vec x}}\frac iD\ket{0,0}_{f_1f^\prime}(\gamma_0\gamma_\mu\hat
Q)_{f^\prime g^\prime}\bra{0,0} {\frac iD} \ket{
-T/2,{\vec y}}_{g^\prime g_1}-\Tr\Bigl(\bra{0,0}{\frac
iD}\,\ket{0,0}\,\gamma_0\gamma_\mu\hat
Q\Bigr)\nonumber\\
&&{}_\times\bra{T/2,{\vec x}}\,{\frac iD}\,\ket{-T/2,{\vec
y}}_{f_1g_1}\Biggr\}
\prod\limits_{i=2}^{N_c}\bra{
T/2,{\vec x}}\,{\frac iD}\,\ket{-T/2,{\vec y}}_{f_ig_i}\,\e^{
iS_{eff}}\,, \label{Eq6a} \end{eqnarray}
where the effective action
\begin{equation}
S_{eff}\,=\,-iN_c\Tr\log D(U)\,.
\label{SEFF}\end{equation}
Since we treat the meson fields classically ({\it i.e.} at
the 0--loop level) and also they are constrained on the chiral circle
(\ref{CHIRCIR}), the  only non-trivial part of the action is the
fermionic part $\Tr\log D(U)$. In eq.(\ref{Eq6a}) in a natural way the
current matrix element is split in a valence -- the first
term, and a Dirac sea contribution -- the second one (see the diagrams on the
l.h.s. of Fig.\ref{Figr1} a) and b)). Since the model is not renormalizable,
in principle, the contribution of the Dirac sea includes divergences and
should be regularized. The second part of this section will be devoted to
the problem of regularization.

In eq.(\ref{Eq6a}) we are left with the integral over the meson fields $U$.
In leading order in $N_c$ it can be done in a saddle point approximation.
To that end we look for a stationary localized meson configuration (soliton)
of hedgehog structure
\begin{equation}
\bar U(x)\,=\,\e^{i\vec{\tau}\cdot\hat{\vec x}\,P(x)}\, ,
\label{HEDGEHOG}\end{equation}
which minimizes the effective action. As it was mentioned before, the
hedgehog soliton $\bar U(x)$ does not preserve the spin and isospin.
As a next step, making use of the rotational zero modes we
quantize it. Since the fluctuations which correspond to the zero
modes are not small, they have to be treated ``exactly'' in the meaning of
path integral. It can be done assuming a rotating meson hedgehog
fields of the form
\begin{equation}
U({\vec x},t)=R(t)\,\bar U({\vec x})\, R^+(t)\, ,
\label{ROTHEDG}\end{equation}
with $R(t)$ being a time-dependent rotation SU(2) matrix in the isospin
space and
$$
\,R^{\dagger} R = \hat 1\,.
$$

It is easy to see that for such an Ansatz one can
transform the effective action
\begin{equation}
\Tr\log D(U)=\Tr\log(D(\bar U)-\Omega)
\label{ROTEFA}\end{equation}
in order to separate the angular velocity matrix:
\begin{equation}
\Omega=-iR^+(t)\dot R(t)=\frac 12\Omega_a\tau_a\,,
\label{OMEGA}\end{equation}
from $D(\bar U)$.
Similar to the effective action the quark propagator in
the background meson fields $U$ can be rewritten as
\begin{equation}
\bra{T/2,{\vec x}}\,\frac i{D(U)}\, \ket{-T/2,{\vec
y}}\,=\,R(T/2)\,\bra{T/2,{\vec
x}}\,\frac i{D(\bar U)-\Omega}\, \ket{-T/2,{\vec y}}\,R^+(-T/2)\,.
\label{Eq10b}\end{equation}
Here, the operator $D(\bar U)-\Omega$ corresponds to the body-fixed
frame of the soliton in which the quark fields are transformed as
\begin{equation}
\Psi\longrightarrow R(t)\Psi\qquad \hbox{and}\qquad
\Psi^\dagger\longrightarrow \Psi^\dagger R^\dagger(t).
\nonumber\end{equation}
Since (as can be seen below from the canonical quantization rules)
$\Omega\sim \frac 1{N_c}$ one can consider $\Omega$ as perturbation and
evaluate any observable as a perturbation series in $\Omega$ which is
actually an expansion in $\frac 1{N_c}$.

In this scheme the matrix element (\ref{Eq2}) of the
current can be written as
\begin{eqnarray}
&&\langle N({\vec p\prime})|\,J_\mu (0)\,\ket{N({\vec p})}\,=\,\frac 1Z\,
N_c\,\int d^3x\,d^3y\,d^3z\,\e^{i{\vec p^\prime\cdot
\vec x}}\,\e^{-i{\vec p\cdot\vec y}}\,\e^{-i({\vec p}^\prime-\vec
p)\cdot\vec z}\nonumber\\
&&{}_\times\int{\cal D}R\
D^{(J)^*}_{-T_3J_3}[R(T/2)]D^{(J)}_{-T_3J_3}[R(-T/2)]
\,\bra{T/2,{\vec x}}\frac i{D-\Omega}\,\ket{-T/2,{\vec y}}^{N_c-1}\nonumber\\
&&{}_\times\Biggl\{\bra{T/2,{\vec x}}\frac
i{D-\Omega}\,\ket{0,{\vec z}}\,R^+(0)
\gamma_0\gamma_\mu\hat Q R(0)\,\bra{0,{\vec z}}\,\frac i{D-\Omega}\,
\ket{-T/2,{\vec y}}\nonumber\\
&&\quad-\Tr\Bigl(\bra{0,{\vec z}}\frac i{D-\Omega}\,\ket{0,{\vec
z}}\,R^+(0)\gamma_0\gamma_\mu\hat Q
R(0)\Bigr)\,\bra{T/2,{\vec x}}\,\frac i{D-\Omega}\, \ket{-T/2,{\vec
y}}\Biggr\}\nonumber\\
&&\qquad\qquad\qquad\qquad\qquad\qquad{}_\times\e^{
N_c\Tr\log(D-\Omega)}\,. \label{Eq12}\end{eqnarray}
Here, the finite rotation matrix (Wigner D-function)
$D^{(J)}_{-T_3J_3}$, which carries the spin
and isospin quantum numbers of the nucleon, appears due to the rotations
$R(t)$ of the valence quark propagators in eq.(\ref{Eq10b}) correlated by
the $\Gamma^{\{g\}}_{JJ_3,TT_3}$ matrices. The D-functions represent the
collective part of the nucleon wave function in the semiclassical
quantization scheme.
The integral over ${\vec z}$ is due to the translational zero modes treated
in the leading order.

Now we are ready to make an expansion in
$\Omega$. For the effective
action it yields~\cite{Diakonov89,Reinhardt89} up to the second order in
$\Omega$:
\begin{equation}
S_{eff}\approx N_c\Tr\log{D}+\frac {\Theta}2\int
dt\Omega_c^2. \label{Eq14} \end{equation}
Here $\Theta$ is the moment of inertia. However, this quantity
includes a divergent contribution from the Dirac sea which should be
regularized. Later we will come back to this point. A derivation of the
moment of inertia including regularization can be found in
ref.~\cite{Reinhardt89} and it has been calculated numerically in
ref.~\cite{Goeke91,Wakamatsu91}.  The first term in eq.(\ref{Eq14})
will be absorbed in $Z$ whereas the second one drives the evolution
in the space of matrix $R$. Expanding the quark
propagator
\begin{equation}
\frac 1{D-\Omega}\longrightarrow\frac 1D +\frac 1D\ \Omega\ \frac 1D+...
\label{OPEXP}\end{equation}
we can separate the zero order ($\sim N_c^0$) and the linear
order ($\sim {\frac 1{N_c}}$) corrections in $\Omega$.
The expansion in  $\Omega$ is illustrated in Fig.\ref{Figr1} a) and  b) for
the valence contribution and for the Dirac sea one, respectively.

After the expansion in $\Omega$ we should deal with the path integral
over $R$. In the case of the zero order term we have
\begin{eqnarray}
\int\d R_1\d R_2\,D^J_{-T_3J_3}(R_1)\,D^{J*}_{-T_3J_3}(R_2)\,
\int\limits_{R_1=R(-T/2)}^{R_2=R(T/2)}{\cal D}R&&\Bigl(R^+(0)\hat Q
R(0)\Bigr)_{fg}\nonumber\\
&&{}_\times\e^{ i\frac{\Theta}2\int \d t^\prime\Omega_c^2(t^\prime)}\,,
\label{Eq16a} \end{eqnarray}
whereas for the linear in $\Omega$ terms we are left
with a more complicated integral:
\begin{eqnarray}
\int\d R_1\d R_2\,D^J_{-T_3J_3}(R_1)\,D^{J*}_{-T_3J_3}(R_2)\,
\int\limits_{R_1=R(-T/2)}^{R_2=R(T/2)}{\cal
D}R&&\Bigl(R^+(0)\hat Q
R(0)\Bigr)_{fg}\,\Omega_c(t)\,\nonumber\\
&&{}_\times\e^{i\frac{\Theta}2\int \d
t^\prime\Omega_c^2(t^\prime)}. \label{Eq16} \end{eqnarray}
For the isoscalar part ${1\over 6} \hat 1$ of the charge matrix $\hat Q$ the
integrals (\ref{Eq16a}) and (\ref{Eq16}) reduce simply to the normalization
of the finite rotation matrix $D^{(J)}_{-T_3J_3}$. In the case of
the isovector part $\frac {\tau^3} 2$ we use the identity
\begin{equation}
\Bigl(R^+(0)\tau^a R(0)\Bigr)_{fg}=\frac 12\,\Tr\Bigl(R^+(0)\tau^a
R(0)\tau^b\Bigr)\,(\tau^b)_{fg} \label{IDENT}\end{equation}
in order to separate the $R(t)$ dependent part of the current which
does not carry flavor indices $f,g$. The path integrals (\ref{Eq16a}) and
(\ref{Eq16}) can be taken rigorously~\cite{Diakonov89}. We essentially made
use of the basic feature of the path integral:
$$
\int\limits_{q_1=q(T_1)}^{q_2=q(T_2)}{\cal D}q\
F_1(q(t_1))\cdots F_n(q(t_n))\,\e^{iS}=\bra{q_2,T_2}\,{\cal T}\{\hat
F_1(q(t_1))\cdots\hat F_n(q(t_n)\}\,\ket{q_1,T_1},
$$
namely that the path integral, which contains $F_i(q(t_1))$, can be
equivalently written
as a matrix element of the time ordered product ${\cal T}$ of the
corresponding operators $\hat F_i(q(t_1))$ . In our case
we have the well-known canonical quantization rule
\begin{equation}
\Omega_c\longrightarrow \frac {\hat J_c}{\Theta}\,,
\label{CANON}\end{equation}
where $\hat J_a$ is the spin operator. For the zero order the collective
path integral (\ref{Eq16a}) is reduced to an ordinary integral of three
Wigner D-functions whereas the final result for (\ref{Eq16}) is a time
ordered product:
\begin{equation}
\vartheta(-t)D_{ab}(R(0))J_c+\vartheta(t)J_c D_{ab}(R(0)),
\label{Eq19} \end{equation}
sandwiched between the nucleon rotational wave functions.
Now using the standard spectral representation of the quark
propagator
\begin{eqnarray}
\langle t^\prime,\vec x^\prime|{\frac 1D}|t,\vec
x\rangle\,&=&\,-i\vartheta(t^\prime-t)\,\sum\limits_{\epsilon_n>0}\e^{
-i\,\epsilon_n(t^\prime-t)}\,\Phi_n(\vec
x^\prime)\,\Phi_n^\dagger(\vec x)\nonumber\\
&&\qquad+i\vartheta(t-t^\prime)\,\sum\limits_{\epsilon_n\le 0}\e^{
-i\,\epsilon_n(t^\prime-t)}\,\Phi_n(\vec
x^\prime)\,\Phi_n^\dagger(\vec x)\,,
\label{SPECTR} \end{eqnarray}
it is straightforward to evaluate the matrix element of the current
$J_\mu$ -- eq.(\ref{Eq12}).
Here $\Phi_\alpha$ and $\epsilon_n$ are the eigenfunctions and
eigenvalues of the single-particle hamiltonian $h$:
\begin{equation}
\Bigl(\frac {\vec{\alpha}\cdot\vec{\nabla}}i+\beta M
U^{\gamma_5}+m_0\beta\Bigr)\Phi_n\,=\,\epsilon_n\Phi_n\,.
\label{EIGEN}\end{equation}

Now we come to the problem of regularization.
Since the NJL model is a non-renormalizable theory,
a regularization scheme of an appropriate cut-off $\Lambda$ is needed to
make the theory finite. In our case it concerns the divergencies
in the Dirac sea contribution to the matrix element of the electromagnetic
current $J_{\mu}$ - diagrams in Fig.\ref{Figr1} b). In Minkowski space-time
it can be equivalently written as:
\begin{eqnarray}
\langle N({\vec p\prime})|&&J_\mu (0)\,\ket{N({\vec p})}_{sea}\equiv
i\,{1\over Z} \,\int \d^3z\,\e^{-i({\vec p}^\prime-\vec p)\cdot\vec z}
\int{\cal D}R\, D^{(J)^*}_{-T_3J_3}[R(T/2)]\nonumber\\
&&{}_\times D^{(J)}_{-T_3J_3}[R(-T/2)]\,
\frac {\delta } {\delta A^{\mu}(0,\vec
z)}\,\e^{ iS_{eff}[\Omega,A^{\mu},R]}\,\Bigg\vert_{A^{\mu}=0}
\,.  \label{CURR} \end{eqnarray}
Here in the effective action $S_{eff}$ (\ref{SEFF})
we include an explicit electromagnetic
coupling in a minimal way:
\begin{equation}
D(\bar U)=i\partial_t\,-\,h(\bar U)\,-\,\Omega\,-\,\hat Q\gamma_0\gamma_\mu
A^\mu\,, \label{DIRACA}\end{equation}
where $A^\mu$ is an external electromagnetic field.

We will work in Euclidean space-time where we
use the metric tensor's signature transformed from $(+ - - -)$ to
$(- - - -)$, i.e. the Euclidean metric tensor $g_{\mu\nu}^{E}\equiv
-\delta_{\mu\nu}$ and the general formulae to perform the transformation
from Minkowski to Euclidean space are:
\begin{eqnarray}
&\,&A^4_E = i A^0_M\,,  \qquad\qquad\quad\ A^k_E = A^k_M \,,  \nonumber\\
&\,&\gamma_E^4 =  i \gamma^0_M\,\  \qquad\qquad\qquad
 \gamma^k_E = \gamma^k_M\,, \nonumber\\
&\,&\tau\equiv x^4_E = i x^0_M\equiv i t\,, \quad\quad x^k_E =
x^k_M\,,\nonumber\\
&\,&\int \d^4 x_E = i \int \d^4 x_M\,, \quad\ \ \partial_\tau = -i
\partial_t  \,,\nonumber\\
&\,&\{ \gamma^E_{\mu}, \gamma^E_{\nu} \} = -2\delta_{\mu\nu}\,, \qquad
\ \{ \gamma_{\mu}^E, \gamma_5^E \} = 0 \,, \nonumber\\
&\,&( \gamma_{\mu}^E )^{\dagger} = - \gamma_{\mu}^E , \qquad\qquad\ \
 (\gamma_5^E)^{\dagger} = \gamma^E_5 = \gamma_5^M \ . \label{EUCLID}
\end{eqnarray}

The transformation of the electromagnetic field $A_\mu$ under the Wick
rotation (transformation to the Euclidean space-time) is defined in
(\ref{EUCLID}) and in order to preserve the gauge invariance the field
$A_\mu$ should be hermitian
in Euclidean space-time. Then in body-fixed frame the Dirac operator of the
rotating soliton $D(\bar U)$  can be expressed in the form:
\begin{equation}
D(\bar U) =  R^{\dagger} \,D(U)\, R\,
= \partial_\tau + h(\bar U) +
i \Omega + i A_4 \, R^{\dagger} \hat Q R -
i \gamma_4 \gamma_k A_k \, R^{\dagger} \hat Q R \ .
\label{DIRFUL} \end{equation}
Here, the rotation velocity matrix $\Omega$ in the
Euclidean space-time is given by:
\begin{equation}
\Omega\equiv{1\over 2}\tau^a \, \Omega^a = - i\ R^{\dagger}\,\dot{R}\,,
\label{RDEF} \end{equation}
where the derivative $\dot R$ is with respect to the Euclidean time
$\tau$ and as usual, summation over repeated indices is
assumed. Similar to ref.~\cite{Reinhardt89} the angular velocity
$\Omega$ is hermitian in Euclidean space-time.
The collective variable $\Omega$
will be quantized  according to the canonical quantization rule
(\ref{CANON}) which in Euclidean space-time is:
\begin{equation}
\Omega_a \longrightarrow -i{\hat{J}_a\over \Theta}\,.
\label{OMQ} \end{equation}
where $\hat{J}^a$ is the spin operator.

Now we can write down the hermitian conjugate to (\ref{DIRFUL}):
\begin{equation}
D^{\dagger}(\bar U) =
- \partial_\tau + h(\bar U)  - i \Omega - i A_4 \,
R^{\dagger} \hat Q R - i \gamma_4 \gamma_k A_k
\,  R^{\dagger} \hat Q R \,,
\label{DIRCON} \end{equation}

As we can see from (\ref{DIRFUL}), (\ref{DIRCON}), the operator
(\ref{DIRFUL}) is non-hermitian. Hence, in general,
the Euclidean action (\ref{SEFF}) will have the real and imaginary part.
To make clear distinction we write down both parts explicitly:
\begin{equation}
S_{eff}^F\, =\, \Re S_{eff}^F + \Im S_{eff}^F \ , \label{REIM}
\end{equation} where
\begin{equation}
\Re S_{eff}^F \equiv - {1\over2}\,N_c \Tr\log
(D^{\dagger} D)\, , \label{REIMa} \end{equation}
end
\begin{equation}
\Im S_{eff}^F \equiv  {i\over2}\, N_c\Tr\log (D/D^{\dagger})\, ,
\label{REIMb}\end{equation}

The imaginary part is finite and does not need
regularization. In addition, any regularization of this part of the
action would break several important features of the model.
On the other hand, the real part is infinite
and a regularization is necessary. We introduce
the Schwinger proper-time regularization of the action.
In general, for an operator $\hat A\hat B^{-1}$
its proper-time regularized form reads ~\cite{Schwinger51}:
\begin{equation}
\Tr\log\bigl(\hat A \hat B^{-1}\bigr)\Rightarrow -\,\Tr
\,\int^{\infty}_{1/\Lambda^2}
{{\d u}\over u}\Biggl(\e^{ -u{\hat A}}\,-\,\e^{ -u{\hat B}}\Biggr)
\,. \label{PROP} \end{equation}

Applying (\ref{PROP}) to the real part of the action (\ref{REIMa}) we
obtain:
\begin{equation}
\Re S^F_{eff} = {N_c \over 2}\,\Tr\int\limits_{1 \over\Lambda^2}^\infty
\frac {\d u}u \,\Biggl(\e^{
-u\,D^{\dagger}D}\,-\,\e^{
-u\,D_v^{\dagger}D_v}\Biggr)\,,\label{REGACTIO} \end{equation}
The cut-off $\Lambda$ is fixed~\cite{TMeissner89}
in the meson sector to reproduce the physical pion decay constant $f_\pi$.
We also subtract the vacuum contribution where the corresponding
operator:
\begin{equation}
D_v=\partial_\tau+h_v
\label{DVACUUM} \end{equation}
includes the unperturbed (vacuum) hamiltonian $h_0$:
\begin{equation}
h_v=\frac {\vec{\alpha}\cdot\vec{\nabla}}i\,+\,\beta M\,+\,m_0\beta\,.
\label{HVACUUM} \end{equation}

Thus the effective action includes the
regularized real part (\ref{REIMa}) and finite imaginary part (\ref{REIMb}).
In fact, for the contribution from the imaginary part (\ref{REIMb}) one can
use the non-regularized expression (\ref{Eq12}).

Now, our task is to evaluate the contribution (\ref{CURR}) coming from
the regularized real part (\ref{REIMa}). In this case the difficulty lies in
that the expression (\ref{REGACTIO}) contains rather complicated operator
exponent $\ \exp({-u\, D^{\dagger} D})\ $ in the integrand. However, because
afterwards we take derivative over the fields $A_{\mu}$ and set $A_{\mu}=0$,
the only non-zero contribution will come from the part of the integrand
linear in $A_{\mu}$. We shall treat the angular velocity $\Omega \sim
{1\over N_c}$ as a small perturbation that is consistent with the large
$N_c$ limit philosophy behind the NJL model.
Hence, for the electromagnetic current we can neglect terms $\Omega^2$ and
higher. However, as can be seen later the term $\sim \Omega^2$ must be
taken into account to evaluate the moment of inertia (see also
ref.~\cite{Reinhardt89}).

In the next step we expand the integrand in
(\ref{CURR}) in terms of $A_{\mu}$ and $\Omega$. It is useful to separate
\begin{equation}
D^{\dagger} D\,=\, D_0^{\dagger} D_0\, +\,\Omega^2\, +\,  V\,,
\label{DDD}\end{equation}
where
\begin{equation}
D_0^{\dagger} D_0\,=\, -\partial_\tau^2 + h^2\,,
\label{D0D0}\end{equation}
and the sum of all perturbative terms we
denote by $V$:
\begin{equation}
V\,=\,W_0\,+\,W_1\,+\,W_2
\label{VDEF}\end{equation}
with
\begin{eqnarray}
&W_0&\,=\,i\,[h\,,\,\Omega]-i\{\Omega\,,\,\partial_\tau\}\,,\label{W0}\\
&W_1&\,=\,[h\,,\,iA_4R^\dagger\hat
QR]+A_4\{R^\dagger\hat QR\,,\,\Omega\}-i\{A_4R^\dagger\hat
QR\,,\,\partial_\tau\}\,,\label{W1} \end{eqnarray}
and
\begin{equation}
W_2=- i\gamma_4\gamma_k\,A_k\{h\,,\,R^\dagger\hat
QR\}+\gamma_4\gamma_k\,A_k[R^\dagger\hat
QR\,,\,\Omega]-i\gamma_4\gamma_k\,[A_kR^\dagger\hat QR\,,\,\partial
_\tau].
\label{W2}\end{equation}
Here $[\ ,\ ]$ and $\{\ ,\ \}$ denote commutators
and anticommutators, respectively. After that for the operator exponent
$\Tr\exp{(-u\,D^{\dagger} D)}$ in eq.(\ref{REGACTIO}) we apply an
expansion
\begin{eqnarray}
\e^{{\hat A}+{\hat B}} =& e^{\hat A}&\, + \,
\int\limits^1_0 \d\alpha\,\e^{\alpha {\hat A}}\ {\hat B}\
\e^{(1-\alpha) {\hat A}}\nonumber\\
&\quad +& \, \int\limits^1_0 \d\beta\,\int^{1-\beta}_0 \d\alpha\,
\e^{\alpha {\hat A}}\ {\hat B}\
\e^{\beta {\hat A}}\ {\hat B}\
\e^{(1-\alpha-\beta) {\hat A}} + ...
\label{FSD}\end{eqnarray}
There for the operator $\hat A$ we take the part
$-u\,D_0^{\dagger} D_0$, which does not contain $A_{\mu}$ and $\Omega$, and
using the cyclic properties of $\Tr$ we get:
\begin{eqnarray}
\Tr\,\e^{ -u\,D^{\dagger} D}\, &=&\, \Tr\,\e^{ -u\,D_0^{\dagger} D_0}
-\,u\, \Tr\,\biggl(\e^{ -u\, D_0^{\dagger} D_0}\, V\biggr)\,
+\,\frac {u^2}2\,\int\limits^1_0 d\beta\nonumber\\
&\quad&{}_\times\Tr \biggl(\,\e^{ -u\,(1-\beta)\,D_0^{\dagger}
D_0}\,V\, \e^{ -u\beta\, D_0^{\dagger} D_0}\,V\biggr)\,+\cdots\,.
\label{SPDD} \end{eqnarray}

We compute the trace over the Euclidean
time in the $\omega$-space, the Fourier conjugate to $\tau$:
\begin{equation}
\bra{\tau} \ f(\partial_{\tau},...)\ket{\tau}\,=\,
\int\limits_{-\infty}^{+\infty} {d\omega\over 2\pi} \ f(-i\omega,...)
\,. \label{TIMESP} \end{equation}

Applying  the operator expansion (\ref{SPDD}),
there are, in general, two types of terms in (\ref{SPDD}) --
linear and bilinear in $\,V\,$. Obviously, in order to give a non-zero
contribution to eq.(\ref{CURR}) all those terms
should be linear $A_{\mu}$. This implies that only $W_{1(2)}$-terms will
contribute to the first order in (\ref{SPDD}):
\begin{equation}
-\,u\,\Tr\,\biggl(\e^{ -u\, D_0^{\dagger} D_0}\,\frac {\delta
W_{1(2)}}{\delta A_\mu(0,\vec z)}\,\Bigg\vert_{A_{\mu}=0}\biggr)\,
\label{FSD1} \end{equation}
One should also keep in mind that $W_{1(2)}$
includes also terms linear in $\Omega$.

The terms contributing to the second
order in (\ref{FSD}) have a more complicated structure:
\begin{equation}
\Tr \biggl(\,\e^{ -u\,(1-\beta)\,D_0^{\dagger}
D_0}\,W_0\, \e^{ -u\beta\, D_0^{\dagger} D_0}\,\frac {\delta W_{1(2)}}
{\delta A_{\mu}(0)}\,\Bigg\vert_{A_{\mu}=0}\,
 +\,W_0\leftrightarrow \frac {\delta W_{1(2)}}{\delta
A_{\mu}}\biggr)\,. \label{FSD2} \end{equation}
In the case that the integral over $R$ includes
both $\Omega_c$ and $D_{ab}$ we are left with
an additional integral over Euclidean time which includes their
time-ordered product (\ref{Eq16}):
\begin{eqnarray}
\int \d \tau e^{i\,(\omega-\omega^\prime)\,\tau}
&\,[&\vartheta(-\tau)\,D_{ab}\,\hat J_c\,+\,\vartheta(\tau)\,
\hat J_c\,D_{ab}]\,=\,{\cal P}\frac
i{\omega-\omega^\prime}\,[D_{ab}\,,\,\hat J_c]\,\nonumber\\
&+&\,\pi\,\delta(\omega-\omega^\prime)\,\{\hat J_c\,,\,D_{ab}\}\,,
\label{TINT}\end{eqnarray}
where in r.h.s. ${\cal P}$ means a principal value.

To summarize, in the scheme presented above the matrix element of the
current $J_\mu$ includes non-regularized as well as regularized
contributions. In general, there are two types of non-regularized
contributions: the valence parts (for all form factors) and
the sea parts coming from the imaginary part of the action. To compute
those terms we use the non-regularized expression (\ref{Eq12}).
For the contribution from the real part of the action we follow the
expressions (\ref{FSD1})~-~(\ref{TINT}) which include an explicit
proper-time regulator (\ref{PROP}).

To begin with explicit evaluation of the particular current matrix elements
and the corresponding form factors we will sketch briefly the derivation
of the Dirac sea contribution to the moment of inertia within the present
scheme. The valence part which does not need regularization and can be found
in ref.~\cite{Diakonov89}:
\begin{equation}
\Theta_{ab}^{val}\,=\,{N_c \over 2}\,\sum\limits_{n\neq val}\frac
{\langle val |\,\tau_a\,|n \rangle \langle
n|\,\tau_b\,|val\rangle}{\epsilon_n-\epsilon_{val}}\,.
\label{THVAL}\end{equation}
Here $|n \rangle$ and $\epsilon_n$ are the eigenfunctions and eigenenergies
of the hamiltonian $h$ and $val$ stands for the valence level.

The Dirac sea part of the moment of inertia $\Theta$ originates from the
 regularized real part $\Re S_{eff}$ of the effective action
(\ref{REIMa}) with electromagnetic field $A_\mu$ set to zero. The imaginary
part does not contribute. In first order operator expansion (\ref{FSD})
only the $\Omega^2$ term contributes:
\begin{eqnarray}
-\frac {N_c}2\int\limits_{1 \over\Lambda^2}^\infty \d
u\,\Tr\Bigl(\e^{ -u\,D_0^\dagger D_0}\tau_a\tau_b\Bigr)  =-{N_c \over
4\sqrt{\pi}}\,\sum\limits_{n,m}\,&& \int\limits_{1
\over\Lambda^2}^\infty \frac{\d u}{\sqrt{u}} \frac
{\e^{-u\,\epsilon_n^2}\,+\,\e^{-u\,\epsilon_m^2}}2\nonumber\\
&&{}_\times\langle n |\,\tau_a\,|m \rangle \langle m|\,\tau_b\,|n\rangle\,,
\label{THFSD1a} \end{eqnarray}
where for convenience we introduce the complete set $|m \rangle \langle m|$
and symmetrize with respect to $n\leftrightarrow m$. The second order term
in expansion (\ref{FSD}) includes only $W_0$:
\begin{equation}
{N_c \over 4}\,\int\limits_{1 \over\Lambda^2}^\infty \d u\, u
\int\limits_0^1\d\beta
\,\Tr\biggl(\,\e^{ -u\,(1-\beta)\,D_0^{\dagger}
D_0}\frac {\delta W_0}{\delta \Omega_a} \e^{ -u\beta\,
D_0^{\dagger} D_0}\frac {\delta W_0}{\delta \Omega_b}
+\,a\leftrightarrow b\,\biggr)\,.
\label{THFSD2} \end{equation}
It is easy to be seen that the above expression is symmetric in
$\,a\leftrightarrow b\,$. After some straightforward calculations one gets:
\begin{equation}
{N_c \over 2} \sum\limits_{n,m} \int\limits_{1
\over\Lambda^2}^\infty
\frac{\d u}{\sqrt{u}} \,\frac
{\e^{-u\,\epsilon_n^2}-\e^{-u\,\epsilon_m^2}}{\epsilon_m^2-\epsilon_n^2}
\,\biggl[\frac 1u+\frac 12\,(\epsilon_m-\epsilon_n)^2\biggr]\nonumber\\
\langle m |\,\tau_a\,|n \rangle \langle n|,\tau_b\,|m\rangle\,.
\label{THFSD2a} \end{equation}

Combining (\ref{THFSD1a}) and (\ref{THFSD2a}) the final result is:
\begin{equation}
\Theta_{ab}^{sea}\,=\,{N_c \over
2}\,\sum\limits_{n\neq m}\,\langle n |\,\tau_a\,|m \rangle \langle
m|\,\tau_b\,|n\rangle \,{\cal R}_\Theta^\Lambda(\epsilon_m\,,\,\epsilon_n).
\label{THFSD} \end{equation}
Here ${\cal R}_\Theta^\Lambda$ is the regularization
function:
\begin{equation}
{\cal R}_\Theta^\Lambda(\epsilon_m\,,\,\epsilon_n)\,= \,{1 \over
4\sqrt{\pi}}
\,\int\limits_{1\over\Lambda^2}^\infty \frac{\d u}{\sqrt{u}} \biggl(\frac
1u\frac {\e^{-u\,\epsilon_n^2}-\e^{-u\,\epsilon_m^2}}
{\epsilon_m^2-\epsilon_n^2} \,-\,\frac
{\epsilon_n\e^{-u\,\epsilon_n^2}\,+\,
\epsilon_m\e^{-u\,\epsilon_m^2}}{\epsilon_m\,+\,\epsilon_n}\biggr)\,,
\label{THREG} \end{equation}
which coincides with ref.~\cite{Reinhardt89}. It is easy to see that the
moment of inertia is diagonal:
\begin{equation}
\Theta_{ab}\equiv \delta_{ab}\Theta\,.
\end{equation}

\section{Computing the form factors}

The nucleon electromagnetic Sachs form factors are related to the matrix
element of the electromagnetic current by:
\begin{eqnarray}
&\langle N(p^\prime)\vert J_0(0)\vert N(p)\rangle &= G_E(q^2)
\ , \label{FFDEFa} \\
&\langle N(p^\prime)\vert \vec J_i(0)\vert N(p)\rangle &=
{1\over 2{\cal M}_N} \ G_M(q^2) \ i \
\, \langle N \vert (\vec\sigma\times\vec q) \vert N \rangle
\ , \label{FFDEFb} \end{eqnarray}
where $\vert N(p) \rangle$ is the nucleon state of a four momentum $p$ and
proper spin and isospin quantum numbers. The four momentum transfer
$q=p^\prime-p$. The electromagnetic current $J_{\mu}(x)$ is in the
Minkowski space-time and ${\cal M}_N$ is the nucleon mass.
The expressions for the matrix element of the current have been
derived in the previous Section.

The isoscalar and isovector parts of the form factors are defined by:
\begin{equation}
G_{E(M)} \equiv {1\over 2}\ G_{E(M)}^{T=0} + \hat T_3 \
G^{T=1}_{E(M)}
\ . \label{FFSV} \end{equation}

The isoscalar form factors (electric and magnetic) come from the
scalar part of the quark current matrix $\hat Q$ ($\sim \hat 1$)
while the isovector form factors is from the triplet part
($\sim \tau^3$). Since the imaginary part of the effective action in
Euclidean space-time (\ref{REIM}) is due to the isoscalar part of the
charge matrix $\hat Q$ one should expect that for both isoscalar (electric
and magnetic) form factors the Dirac sea contribution originates from the
imaginary part of the effective action. Indeed, as can be checked directly
the contribution from the regularized real part of the
action to both isoscalar form factors is exactly zero.
This means that for the calculation of these form factors the
non-regularized expression (\ref{Eq12}) can be used.

We start with the electric scalar form factor. Since only the isoscalar part
of the operator $\hat Q$ contributes ($\hat J_0\,=\,\frac16\hat 1$) the
rotational matrices $R(t)$  vanish from the action. Hence, the $D{\cal R}$
integration in (\ref{Eq12}) becomes trivial. It can be seen easily from
(\ref{FSD1}) and (\ref{FSD2}) using the explicit form of $W_0$ and $W_1$
that because of symmetry considerations there is no contribution from the
real part of the action and the Dirac sea contribution comes only from the
imaginary part (\ref{REIMb}). We start with the non-regularized
eq.(\ref{Eq12}), expand the effective action and the quark
propagator in $\Omega$, (\ref{Eq14}) -(\ref{OPEXP}), and insert the spectral
representation (\ref{SPECTR}). In this case the contribution from
the terms linear in $\Omega$ is exactly zero. Finally we arrive at:
\begin{eqnarray}
G_E^{T=0} (q^2)\, =\, \int \d^3 &z\,& e^{ i \vec q \vec z}
\ {N_c\over 3}
\ \Big\lbrace \Phi_{val}^{\dagger}(\vec z) \,
\Phi_{val}(\vec z)\nonumber\\
&\,&\qquad\qquad -\,{1\over 2}\ \sum\limits_{n}\,{\rm sign}(\epsilon_n)
\,\Phi_n^{\dagger}(\vec z) \,\Phi_n(\vec z)
\,\Big\rbrace \ ,
\label{ELSCAL} \end{eqnarray}
The first term in eq.(\ref{ELSCAL}) comes from the valence
quarks and the second term is due to the quark vacuum polarization.

At $q^2=0$ the isoscalar electric form factor reproduces the baryon
number $B=1$: it easy to see that the contribution from the sea vanishes
exactly whereas the valence term gives one (normalization of $\Phi_{val}$).

As the next step, we compute the electric isovector form factor.
Using the formulae (\ref{Eq12})-(\ref{SPECTR}) the evaluation of the
valence part is straightforward (see ref.~\cite{Diakonov89}) and one gets
\begin{eqnarray}
\bra{N(p^\prime)}\Psi^\dagger\frac
{\tau^3}2\Psi\ket{N(p)}_{val}=-\frac{N_c}{2\Theta}&&
\int \d z \e^{-i\vec q\cdot\vec z} \sum\limits_{m\neq val} \frac
{\langle val |\tau_c|m \rangle(\Phi_m^\dagger(\vec
x)\tau_3\Phi_{val}(\vec
x))}{\epsilon_m-\epsilon_{val}}\nonumber\\
&&{}_\times\bra{1/2,J_3T_3}\{\hat J_c \,,\,D_{3b}\}\ket{1/2,J_3T_3}\,.
\label{GEV}\end{eqnarray}
The anticommutator $\,\{\hat J_c\,,\,D_{3b}\}\,$ appears due to the
symmetry of the $\tau$ matrix element product under $\ n \leftrightarrow m\
$.

The contribution of the Dirac sea comes from the real part of the
effective action and a regularization is necessary. Contributions to the
first order in expansion (\ref{SPDD}) come only from
the term of $W_1$ containing both $A_4$ and $\Omega$:
$\{A_4R^\dagger\hat \tau^3R\,,\,\Omega\}\,$:
\begin{eqnarray}
&&\Tr\Bigr(\e^{-u\,D^\dagger_0D_0}\frac {\delta W_1}{\delta
A_4(0,\vec z)}\,\Bigg\vert_{A_4=0}\Bigl)
=\,{N_c \over 4\sqrt{\pi}}
\sum\limits_{n,m} \int\limits_{1
\over\Lambda^2}^\infty \frac{\d u}{\sqrt{u}} \frac
{\e^{-u\epsilon_n^2}+\e^{-u\epsilon_m^2}}2\nonumber\\
&&{}_\times\langle n |\,\tau_c\,|m \rangle \Bigl(\Phi_m^\dagger(\vec
z)\tau^3\Phi_n(\vec z)\,\Bigr)\bra{1/2,J_3T_3}\{\hat J_c
\,,\,D_{3b}\}\ket{1/2,J_3T_3}\,. \label{GEFSD1} \end{eqnarray}
The other two terms in $W_1$, $i[h\,,\,A_4R^\dagger\frac
{\tau^3}2R]$ and $\,-i\{A_4R^\dagger\frac {\tau^3}2R\,,\,\partial_\tau\}\,$,
give no contribution.

The second order in (\ref{SPDD}) includes terms from $W_0$
and $W_1$: \begin{eqnarray}
-i\,{N_c \over 4} \int\limits_0^1\d\beta\,(1-\beta) \int\limits_{1
\over\Lambda^2}^\infty
\d u&& u\,\Tr\biggl(\,\e^{ -u\,D_0^{\dagger}
D_0}\, W_0\, \e^{ -u\beta\, D_0^{\dagger}
D_0}\,\frac {\delta W_1}{\delta A_4(0,\vec
z)}\,\Bigg\vert_{A_4=0}\nonumber\\
&&+\, W_0\leftrightarrow \frac {\delta W_1}{\delta A_4}
\biggr)\, \label{GEFSD2} \end{eqnarray}
Because of the symmetry of the $\tau$ matrix element product under $\
n~\leftrightarrow~m\ $ only the symmetric part of the r.h.s. of the
integral (\ref{TINT}) survives. After integrating over $\omega$ and
$\beta$ (the integrals are the same like in the case of the moment of
inertia) one obtains:
\begin{eqnarray}
\,-\,{N_c \over 4\sqrt{\pi}\Theta}&& \sum\limits_{n,m} \int\limits_{1
\over\Lambda^2}^\infty
\frac{\d u}{\sqrt{u}}\,\frac
{\e^{-u\,\epsilon_n^2}-\e^{-u\,\epsilon_m^2}}{\epsilon_m^2-\epsilon_n^2}
\,\bigl[\frac 1u\,+\,\frac 12\,(\epsilon_m-\epsilon_n)^2\bigr]
\langle n |\,\tau_c\,|m \rangle\nonumber\\
&&{}_\times\Bigl(\Phi_m^{\dagger}(\vec z)\,\tau_3\,\Phi_n(\vec z)\,\Bigr)
\,\bra{1/2,J_3T_3}\{\hat J_c\,,\,D_{3b}\}\ket{1/2,J_3T_3}\,.
\label{GEFSD2a} \end{eqnarray}

Finally, using the quantization rule (\ref{OMQ}) the collective
matrix element can be easily calculated:
\begin{equation}
\bra{1/2,J_3^\prime T_3^\prime}\{\hat
J_c\,,\,D_{ab}\}\ket{1/2,J_3T_3}=-\,\frac
13\, \delta_{cb}\,(\tau^a)_{{}_{T_3^\prime T_3^{}}}\,\delta_{{}_{J_3^\prime
J_3^{}}}\,. \label{ACOMM} \end{equation}
Combining (\ref{GEV}) - (\ref{GEFSD2a}) we arrive at the following
expression for the electric isovector form factor in Minkowski space-time:
\begin{eqnarray}
G_E^{T=1} (q^2)& = &{N_c\over 6\Theta}\, \int d^3 x\ e^{-i \vec q \vec z}
\ \Bigg\lbrace\ \sum_{n}
{\Bigl(\Phi^{\dagger}_{val}(\vec z)\tau^a\Phi_n(\vec z)\Bigr)\,
\bra{n}\tau^a\ket{val}  \over \epsilon_n -
\epsilon_{val}}\nonumber\\
&\,&\qquad+ \sum_{m,n} {\cal R}_\Theta^\Lambda(\epsilon_m,\epsilon_n)
\Bigl(\Phi_n^{\dagger}(\vec z)\tau^a \Phi_m(\vec z)\Bigr)\,
\bra{m}\tau^a\ket{n} \Bigg\rbrace \ ,
\label{ELVEC} \end{eqnarray}
where the regularization function ${\cal R}_\Theta^\Lambda$ is exactly the
same as in the case of moment of inertia. Hence, formula (\ref{ELVEC}) for
$q^2=0$ is equivalent to the quotient of the moment of inertia by itself:
\begin{equation}
G^{T=1}_E\Bigl(q^2=0\Bigr) = 1\,,
\end{equation}
as it should be.

Now we proceed to compute the magnetic form factors. We start with
the isoscalar one. In this case, the quark charge matrix $\hat Q$
contributes {\it via} the isoscalar term, $\sim {1\over 6}\,\,\hat 1$,
diagonal in the isospin space, the rotation matrices $R(t)$ cancel each
other and the current is $\sim \frac 16\gamma_0\gamma_k$. Similar to the
isoscalar electric form factor, the Dirac sea contribution from the real
part of the action is zero: the term from the first order in (\ref{SPDD}) is
proportional to $\,\bra{n}\gamma_0\gamma_k\ket{n}\,=\,0\,$ and in the second
order the terms are odd in $\omega$ and after integration vanish.
Hence, the Dirac sea contribution comes solely from the imaginary part of
the action and the current matrix element can be
calculated directly from (\ref{Eq12}). However, since the current does not
include rotation matrices $R(t)$ the first (zero order in $\Omega$) term
from (\ref{OPEXP}) gives no contributions. In the leading order
(linear in $\Omega$) the collective part includes only $\Omega$:
\begin{eqnarray}
\bra{N(p^\prime)}\Psi^\dagger\gamma_0\gamma_k\Psi\ket{N(p)}=
-\frac{N_c}{2\Theta} \int \d^3 z \e^{ -i\,\vec q\cdot\vec z}
&&\sum\limits_{n > val\atop m\leq val}{(\Phi_m^{\dagger}(\vec
z)\gamma_0\gamma_k\Phi_ n(\vec z))\,\bra{n}\tau^c\ket{m}\over
\epsilon_n- \epsilon_m} \nonumber\\
&&{}_\times\bra{1/2,J_3T_3}\hat J_c\ket{1/2,J_3T_3}\,.
\label{GMM} \end{eqnarray}
{}From (\ref{FFDEFb}) and (\ref{GMM}) resolving the spin structure we get for
the magnetic isoscalar form factor the following expression:
\begin{equation}
G_M^{T=0} (q^2) = \,i\, {N_c \over6 \Theta}\,{\cal M}_N\,\int\d^3
z\e^{ -i\vec
q \vec z}\,\epsilon_{kja} \,  {q_j\over q^2}
\sum_{n > val\atop m\leq
val} {\Phi_m^{\dagger}(\vec z)\gamma_0\gamma_k\Phi_n(\vec z)\,
\bra{n}\tau^a\ket{m}  \over \epsilon_n - \epsilon_m},
\label{MAGSC} \end{equation}
where, as usually, summation over repeated indices $k, j, a$
is assumed.

For the magnetic isovector form factor the current $J_\mu$ includes
the isotriplet part $\tau^3/2$ of the quark charge matrix
$\hat Q$. Because of this we cannot get rid of the rotation
matrix $R(t)$ in the action. The valence part is obtained from
eq.(\ref{Eq12}). Expanding the quark propagator in $\Omega$ and using the
 spectral representation (\ref{SPECTR}) it is straightforward to
calculate the contributions from the first as well as from the second order
terms in expansion (\ref{OPEXP}). For the first order (zero in $\Omega$)
there is no problem
with time-ordering and the collective integral (\ref{Eq16}) is reduced to an
ordinary integral of three Wigner D-functions:
\begin{equation}
\bra{1/2,J_3^\prime T_3^\prime}\,D_{ab}\,\ket{1/2,J_3T_3}\,=\,-\frac
13\,(\tau^a)_{{}_{T_3^\prime T_3^{}}}\,(\tau^b)_{{}_{J_3^\prime J_3^{}}}
\, \label{WDF} \end{equation}
In the second order (linear in $\Omega$) we have a time-ordered product
(\ref{Eq19}). In contrast to the isoscalar magnetic form factor the quark
matrix element\newline
$\,\bra{n}\gamma_0\gamma_k\tau^3\ket{m}\bra{m}\tau^a\ket{n}\,$
is asymmetric under $\ n \leftrightarrow m\ $ and we end up with a
commutator in the collective part:
\begin{equation}
\bra{1/2,J_3^\prime T_3^\prime}[\hat J_c,D_{ab}]\ket{1/2,J_3T_3}\,=\,i
\varepsilon^{cbd}\,\bra{1/2,J_3^\prime T_3^\prime}D_{ad}\ket{1/2,J_3T_3}
\label{Eq21} \end{equation}
Summing up the first and the second order terms we get for the valence part:
\begin{eqnarray}
&&\bra{N(p^\prime)}\Psi^\dagger\gamma_0\gamma_k\frac
{\tau^3}2\Psi\ket{N(p)}_{val}\,=\,
N_c\int \d^3 z \e^{-i\,\vec q\cdot\vec z}\nonumber\\
&&{}_\times\Biggl\{\Bigl(\Phi^\dagger_{val}({\vec
z})\,\gamma_0\gamma_k\tau^a\,\Phi_{val}({\vec
z})\Bigr)\bra{1/2,J_3T_3}D_{3b}\ket{1/2,J_3T_3}
+\frac 1{2\Theta}\sum\limits_{n\neq val}\sign({\epsilon_n})\nonumber\\
&&{}_\times\frac {\Bigl(\Phi^\dagger_{val}({\vec
z})\gamma_0\gamma_k\tau^b\Phi_n({\vec
z})\Bigr)\bra{n}\tau^c\ket{val}}{\epsilon_n -
\epsilon_{val}}\bra{1/2,J_3T_3}[\hat J_c
\,,\,D_{3b}]\ket{1/2,J_3T_3}\Biggr\}\,. \label{Eq20}\end{eqnarray}
The Dirac sea part originates from the regularized real part of the
effective action (\ref{REIM}) (in Euclidean space-time).
In the first order in (\ref{SPDD}) the only term from $W_2$ (\ref{W2}),
which gives
non-zero contribution, is $\,i\gamma_4\gamma_k\,\{h\,,\,R^\dagger\hat QR\}$.
The other two, $\,i\gamma_4\gamma_k\,[R^\dagger\hat QR\,,\,\partial_\tau]\,$
and
$-\gamma_4\gamma_k\,[R^\dagger\hat QR\,,\,\Omega]\,$ simply vanish. Using
(\ref{WDF}) one gets
\begin{eqnarray}
N_c\int\limits_{\frac 1{\Lambda^2}}^\infty \d u &&
\Tr\biggl(\e^{-u\,D^\dagger_0
D_0} \frac {\delta W_2}{\delta A_k(0,\vec
z)}\,\Bigg\vert_{A_k=0}\biggr)\,= \,
-N_c \sum\limits_n {\cal R}_{M1}^\Lambda(\epsilon_n)\nonumber\\
&&{}_\times\Bigl(\Phi^\dagger_n({\vec z}) \gamma_4 \gamma_k \tau^3
\Phi_n({\vec z})\Bigr)\bra{1/2,J_3T_3}D_{3b}\ket{1/2,J_3T_3}\,,
\label{MVCURp} \end{eqnarray}
with ${\cal R}_{M1}^\Lambda$ being the corresponding regulator:
\begin{equation}
{\cal R}_{M1}^\Lambda(\epsilon_n)\,=\,{1\over 2
\sqrt{\pi}}\int_{1\over\Lambda^2}^{\infty}\frac
{\d u}{\sqrt{u}}\epsilon_n \,  \e^{-u\epsilon_n^2}\,.  \label{MVREG1}
\end{equation}
To the second order
\begin{eqnarray}
-{N_c \over 4} \int\limits_0^1\d\beta\int\limits_{1\over\Lambda^2}^\infty
\d u u&&\Tr\biggl(\,\e^{ -u\,(1-\beta)\,D_0^{\dagger}
D_0}\, W_0\, \e^{ -u\beta\, D_0^{\dagger}
D_0}\,\frac {\delta W_1}{\delta A_k(0,\vec
z)}\,\Bigg\vert_{A_k=0}\nonumber\\
&&+\, W_0\leftrightarrow \frac {\delta W_1}{\delta A_k}\biggr) \,,
\label{GEFSD2b} \end{eqnarray}
contribute $\,i\gamma_4\gamma_k\,\{h\,,\,R^\dagger\hat QR\}$ and
$\,i\gamma_4\gamma_k\,[R^\dagger\hat QR\,,\,\partial_\tau]\,$. Using the
asymmetry of the quark matrix element
$\,\bra{n}\gamma_4\gamma_k\tau^3\ket{m}\bra{m}\tau^a\ket{n}\,$
under $\ n~\leftrightarrow~m\ $ one can show that the symmetric part
of the r.h.s. of the integral (\ref{TINT}) vanishes.
Only the asymmetric part survives:
\begin{eqnarray}
&&\frac {N_c}{2\Theta}\int \d\beta\int \d u\, u \,\int \frac
{\d\omega}{2\pi}\,\int \frac {\d\omega^\prime}{2\pi}\e^{
-u\,(1-\beta)\,(\omega^2+\epsilon_m^2)}\,\e^{
-u\,\beta\,({\omega^\prime}^2+\epsilon_n^2)}
\frac 1{\omega-\omega^\prime}\,(\omega\epsilon_m\,+
\,\omega^\prime\epsilon_n)\nonumber\\
&&{}_\times\Bigl(\Phi^\dagger_m({\vec
z})\gamma_4\gamma_k\tau^b\Phi_n({\vec
z})\Bigr)\bra{n}\tau^c\ket{m}
\bra{1/2,J_3T_3}[\hat J_c ,D_{ab}]\ket{1/2,J_3T_3}
\label{MVFDS2} \end{eqnarray}
Using (\ref{Eq21}) and performing the integrations over $\omega$ and $u$
we get for the second order contribution in the form:
\begin{equation}
i{N_c\over 9\Theta}  \sum\limits_{n,m}{\cal
R}_{M2}^\Lambda(\epsilon_m,\epsilon_n)\Bigl(\Phi_m^{\dagger}({\vec z})
\gamma_0 \gamma_k \tau_b\Phi_n({\vec z})\Bigr)
\bra{n}\tau^c\ket{m}
\bra{1/2,J_3T_3}[\hat J_c\, ,\,D_{ab}]\ket{1/2,J_3T_3}\,,
\label{MVFDS2a}\end{equation}
with a second regulator ${\cal R}_{M2}^\Lambda$:
\begin{equation}
{\cal
R}_{M2}^\Lambda(\epsilon_m,\epsilon_n)\,=\,\frac1{4\pi}
\int\limits_0^1\,\frac {\d\beta}{\sqrt{(1-\beta)\beta}}\,\frac
{\epsilon_m\,-\,\beta\,(\epsilon_n + \epsilon_m)}{(1-\beta)\epsilon^2_m\,+
\,\beta\epsilon^2_n}\,\e^{ -[(1-\beta)\epsilon^2_m\,+ \,
\beta\epsilon^2_n]/\Lambda^2}\,.
\label{MVREGa} \end{equation}
Summing all obtained contributions (\ref{Eq20})-(\ref{MVFDS2a}) and using
eqs.(\ref{FFDEFb}) and (\ref{FFSV}),
we obtain the final expression for the magnetic isovector
form factor in the following form:
\begin{eqnarray}
G_M^{T=1}(q^2)&& =i{\cal M}_N {N_c\over 3}\int \d^3 z\, \e^{-i \vec
q \vec z}\,\varepsilon_{kja} \, {q_j\over q^2}\nonumber\\
&&{}_\times\bigg\lbrace \Bigl(\Phi_{val}^{\dagger}(\vec
z)\gamma_0\gamma_k\tau^a\Phi_{val}(\vec z)\Bigr)
-\,\sum\limits_{n}\,{\cal R}_1^M(\epsilon_n)\,
\Bigl(\Phi_n^{\dagger}(\vec z)\gamma_0\gamma_k\tau^a
\Phi_n(\vec z)\Bigr) \nonumber\\
&&+\,\frac i{2\Theta}\varepsilon^{abc}\sum\limits_{
n\neq val}\sign({\epsilon_n})\,\frac
{\Bigl(\Phi^\dagger_{val}({\vec z})\gamma_0\gamma_k\tau^b\Phi_n({\vec
z})\Bigr)\bra{n}\tau^c\ket{val}}{\epsilon_n - \epsilon_{val}}\nonumber\\
&&+\, {i\over 2 \Theta} \, \varepsilon^{abc}
\sum\limits_{n,m}\,{\cal R}_{M2}^\Lambda(\epsilon_m,\epsilon_n)
\,\Bigl(\Phi_n^{\dagger}({\vec z}) \gamma_0 \gamma_k \, \tau^b
\,\Phi_m({\vec z})\Bigr) \bra{m} \tau^c  \ket{n}\bigg\rbrace\,.
\label{MAGVEC} \end{eqnarray}

The proton and neutron form factors are defined, respectively, as sum
and difference of the isoscalar and isovector form factors:
\begin{eqnarray}
G^p_{E,M} &= {1\over 2}\ \lbrack \, G^{T=0}_{E,M} + G^{T=1}_{E,M}
\, \rbrack \ , \label{FFPNa} \\
G^n_{E,M} &= {1\over 2}\ \lbrack \, G^{T=0}_{E,M} - G^{T=1}_{E,M}
\, \rbrack \ , \label{FFPNb} \end{eqnarray}
and the numerical results for them are presented in
Section 4.

\section{Numerical results}

To compute observables we use and numerical method of Ripka and Kahana
{}~\cite{Ripka84} for solving the eigenvalue problem in finite quasi--discrete
basis. We consider a spherical box of a large radius
$D$ and the basis is made discrete by imposing a boundary condition at
$r=D$. Also, it is made finite by restricting momenta of the basis states
to be smaller than the numerical cut-off $K_{max}$.
Both quantities have no physical meaning and the results
should not depend on them. The typical values used are
$D \sim 20/M$ and $K_{max} \sim 7 M$.

In addition, all checks concerning the numerical stability of the
solution with respect to varying box size and choice of the
numerical cut-off have been done and the actual calculation is
completely under control.

The actual calculations are done by fixing the parameters in
meson sector in the well known way ~\cite{TMeissner89} to have
$f_{\pi} = 93\ MeV$ and $m_{\pi} = 139.6\ MeV$.
This leaves the constituent quark mass $M$ as the only free
parameter.

The proton and neutron electric and magnetic form factors
are displayed in Figs.\ref{Figr2}-\ref{Figr5}. The theoretical curves
resulting from the model are given for the following four values the
constituent quark mass: $370, 400, 420$ and $450\ MeV$.
The magnetic form factors are normalized to the experimental
values of the corresponding magnetic moments at $q^2=0$.
With one exception of the neutron electric form factor (Fig.\ref{Figr3}),
all other form factors fit to the experimental data quite well. The best fit
is for the constituent quark mass around $400$--$420$ MeV.

\vskip\baselineskip
\hrule
\vskip\baselineskip
\centerline{\bf Figs.\ref{Figr2}-\ref{Figr5}}
\vskip\baselineskip
\hrule
\vskip\baselineskip

As can be seen the only form factor which deviates from the experimental
data is the neutron electric form factor and this requires
some explanation. Obviously, this form factor is the most sensitive
for numerical errors.  According to the formula (\ref{FFPNb})
the form factor has been calculated as a difference of the electric
isoscalar and electric isovector form factors that were the direct
output  of our code. Both form factors were of order $1.00$
and calculated by the code with a high enough accuracy.
However, the resulting neutron form factor has the proper (experimental)
value of order $0.04$, {\it i.e.} about $4\%$ of the value of its
components. This means that a small numerical error for one of the
components can be enhanced by a factor $50$.
Hence, small numerical errors together with the applied approximations
(like the $1/N_c$ approximation behind the model and
neglecting the boson--loop effects as well as higher order fermion
loops) are strongly magnified resulting in a deviation from
the experimental data for momentum transfers above $100\ MeV/c$.

\vskip\baselineskip
\hrule
\vskip\baselineskip
\centerline{\bf Table~\ref{Tabl1}}
\vskip\baselineskip
\hrule
\vskip\baselineskip

 As the next step, one can compute other electromagnetic observables:
the mean squared radii, the magnetic moments and the
nucleon--$\Delta$ splitting. The static nucleon properties, in particular
the charge radii and the magnetic moments can be obtained from the form
factors:
\begin{equation}
\langle r^2 \rangle_{T=0,1} = -{6\over G_E^{T=0,1}}\, {\d G_E^{T=0,1}\over
\d q^2}\, \Bigg\vert_{q^2=0} \ ,
\label{ELRAD} \end{equation}
\begin{equation}
\mu^{T=0,1} = G_M^{T=0,1} (q^2) \Big\vert_{q^2=0} \ .
\label{MAGMOM} \end{equation}

For the quark masses $370, 420$ and $450\
MeV$  and pion mass $m_{\pi}=140 MeV$ the calculated values are
presented in Table~\ref{Tabl1}. The values for the axial vector coupling
constant $g_A$, presented for completeness, are from ref.~\cite{Christov94}.
For comparison, in Table~\ref{Tabl2}
we give the theoretical values of the same quantities but with the physical
pion mass set to zero.

The chiral limit ($m_{\pi}\to 0$) mostly
influences the isovector charge radius. Indeed, as it should be
expected~\cite{Bet72,Adkins83}, in our calculations with zero pion mass the
isovector
charge radius diverges in chiral limit. It is illustrated in Fig.\ref{Figr6},
where the isovector charge radius is plotted {\it vs.} the box size $D$.
The Dirac sea contribution to the radius grows linearly with $D$ and
diverges as $D\to\infty$. Because of this that quantity (and the relative
quantities) is not included in Table~\ref{Tabl2}.
The other observable strongly influenced by the chiral limit
is the neutron electric form factor.
For the $m_{\pi}\to 0$ the discrepancy from the experiment is
by almost a factor two larger than in the case $m_{\pi}\neq 0$.
The other observables differ in the chiral limit by about 30\%.
The comparison of the values in the two tables indicate that
taking the physical pion mass gives us a best fit with
a much better agreement with the experimental data.
To be particular, in chiral limit ($m_{\pi}\,=\,0$
the calculations for the electric observables suggest a
high value ($M\sim 450\ MeV$) while for the magnetic ones indicate
$M\sim 370\ MeV$ which is not the case with physical pion mass. In
addition, in chiral limit we observe much larger contribution from the sea
effects, about 50\% of the total value.

\vskip\baselineskip
\hrule
\vskip\baselineskip
\centerline{\bf Table~\ref{Tabl2}}
\vskip\baselineskip
\hrule
\vskip\baselineskip

The results of Table~\ref{Tabl1}   ($m_{\pi}=140$ MeV) again indicate the
value $\sim 400$--$420$ MeV for the constituent quark mass, in
agreement with the conclusion drawn from the form factor curves.
The same has been suggested earlier ~\cite{Goeke91,Gorski92}, where
smaller number of observables has been taken into account.
With the exception of the neutron electric squared radii, to
which remarks similar to the case of the neutron electric form factor
are applicable, the contribution of the valence quarks is dominant. However,
the contribution of the Dirac sea is non-negligible, it lies within the
range 15 -- 40\%.

\vskip\baselineskip
\hrule
\vskip\baselineskip
\centerline{\bf Fig.\ref{Figr6}}
\vskip\baselineskip
\hrule
\vskip\baselineskip

One can notice, that the numerical results for
the nucleon--$\Delta$ mass splitting ($M_{\Delta}-M_N$),
the mean squared proton, isoscalar and isovector
electric radii and the axial coupling constant ($g_A$),
as well as the proton electric and magnetic and neutron
magnetic form factor, differ from the experimental data by no more then
about $\pm 5\%$. Finally, for the magnetic moments we have got results
20--25\% below their experimental values. Despite of this
underestimation of
both magnetic moments we have found surprisingly good result for the
ratio $\mu_p/\mu_n$ which is far better than in other models.
The agreement with the experimental value is better than
3\% for the constituent quark mass 420 $MeV$.
So good result has been obtained neither in the Skyrme model nor
in the linear sigma model.

In section 3 we have found that the magnetic isovector form factor is the
only one which includes non-zero contributions in both leading ($N_c^0$) and
next to leading order ($1/N_c$) in $\Omega$. However, the numerical
calculations show that the ($1/N_c$) rotational corrections do not
affect the $q^2$-dependence (slope) of the form factor but rather the
value at the origin ($G_M^{T=1}(q^2=0)\,=\,\mu^{T=1}$) which is the
isovector magnetic
moment. It is not surprising since (as can be seen from eq.(\ref{MAGVEC}))
in both leading and next to leading order terms the shape of the wave
functions $\Phi_n$ determines the $q^2$-dependence of the form factor
whereas the value at $q^2=0$ depends on the particular matrix elements
included. In leading order the isovector magnetic moment is
strongly underestimated~\cite{Wakamatsu91}. As has been shown in
ref.~\cite{Christov94} the enhancement for this quantity
due to the $1/N_c$ corrections is of order $(N_c+2)/N_c$ and it improves
considerably the agreement with experiment. However, as can be seen from
Table \ref{Tabl1} it is still below the experimental value by 25 \%.

The isoscalar and isovector electric mean squared radii are
shown in Figs.\ref{Figr7},\ref{Figr8} as functions of the constituent
quark mass. The same plot but for the experimentally measured
quantities: the proton and neutron electric charge radii,
is given in Fig.\ref{Figr9}.
In these plots the valence and sea contributions are explicitly
given (dashed and dash-dotted lines, respectively). As could be
expected,
for the isoscalar electric charge radius the valence part is dominant
(about 85\%), due to the fact that there is no $1/N_c$ rotational
contributions. This is not true for the isovector electric charge
radius, where the sea part contributes to about 45\% of the total value
(Fig.\ref{Figr7}). Also, this effect can be seen from the proton and neutron
charge radii (Fig.\ref{Figr9}) which are linear combinations of the
isoscalar and isovector ones.
For proton the sea contribution is about 30\%. However, for the
neutron charge radius the negative sea part is dominating and the valence
contribution is negligible.

\vskip\baselineskip
\hrule
\vskip\baselineskip
\centerline{\bf Fig.\ref{Figr7} and \ref{Figr8}}
\vskip\baselineskip
\hrule
\vskip\baselineskip

In addition, in Fig.\ref{Figr10} we plot the proton and neutron charge
distribution for the constituent quark  mass $M\,=\,420\,$ MeV. For the
proton we have a positive definite charge distribution completely dominated
by the valence contribution whereas in the case of the neutron the Dirac sea
is dominant. In accordance with the well accepted phenomenological
picture one realizes a positive core coming
from the valence quarks and a long negative tail due to the polarization
of the Dirac sea. Using the gradient expansion the latter can be expressed
in terms of the dynamical pion field -- pion cloud.

For completeness, in Fig.\ref{Figr11} we present also the magnetic moment
density distribution for proton and neutron for the constituent quark
mass $M\,=\,420\,$ MeV. The sea contribution becomes non-negligible for
distances greater than $0.5\ fm$. Due to the relatively large tail
contribution to the magnetic moments the sea contribution to these quantities
is about 30\%.

\vskip\baselineskip
\hrule
\vskip\baselineskip
\centerline{\bf Fig.\ref{Figr9}}
\vskip\baselineskip
\hrule
\vskip\baselineskip

\centerline{\bf 5. Summary and conclusions}

Our numerical results support the view that the chiral quark soliton model
of the Nambu--Jona-Lasinio type offers relatively simple but
quite successful description of some low-energy QCD phenomena and,
in particular, of the electromagnetic properties of nucleons. In parameter
free calculations we have obtained overall good results for the
electromagnetic form factors, the mean squared radii, the magnetic moments
and the nucleon--$\Delta$ splitting.
Using $f_{\pi} = 93\ MeV$ and $m_{\pi} = 139.6\ MeV$ and a
constituent quark mass of $M = 420\ MeV$ the isoscalar and
isovector electric radii are reproduced within 15\%. The magnetic
moments are underestimated by about 25\%, however, their ratio
$\mu_p/\mu_n$ is almost perfectly reproduced. The $q$--dependence
of $G^p_E(q^2)$, $G^p_M(q^2)$ and $G^n_M(q^2)$ is very well reproduced up
to momentum transfer of 1 GeV. The neutron form factor $G^n_E(q^2)$ is by a
factor of two too large for $q^2 > 100\ MeV^2$. One should note, however,
that $G^n_E(q^2)$ is more than an order of magnitude smaller
than $G^p_E(q^2)$  and as such it is extremely sensitive
to both the model approximations and the numerical techniques used.
Here, it turns out that the agreement
is noticeably worse if the chiral limit ($m_{\pi}\to 0$) is
used. This can be easily understood since the pion mass
determines the asymptotic behaviour of the pion field.
Altogether we conclude that the
chiral quark soliton model based on a bosonized NJL type lagrangean is
quite appropriate for the evaluation of nucleonic electromagnetic
properties.

\noindent{\bf Acknowledgement:}

We would like to thank Victor Petrov for numerous helpful discussions.
The project has been partially supported by the BMFT,
DFG and COSY (J\"ulich).
Also, we are greatly indebted for financial support to
the Bulgarian Science Foundation $\Phi$--32
(CVC) and to the Polish Committee for
Scientific Research,  Projects KBN nr 2 P302 157 04
and 2 0091 91 01 (AZG).

\newpage
\begin{figure}
\vskip2cm
\caption{Diagrams corresponding to the expansion in $\Omega$ of the
electromagnetic current matrix element:
a) the valence contribution and b) the Dirac sea contribution.}
\label{Figr1}
\end{figure}
\begin{figure}
\caption{The proton electric form factor for the
momentum transfers below $1\ GeV$.}
\label{Figr2}
\end{figure}
\begin{figure}
\caption{ The neutron electric form factor for the
momentum transfers below $1\ GeV$.}
\label{Figr3}
\end{figure}
\begin{figure}
\caption{ The proton magnetic form factor normalized to the experimental
value of the proton magnetic moment at $q^2=0$ for the momentum transfers
below $1\ GeV$. The normalization factor can be extracted from Table 1.}
\label{Figr4}
\end{figure}
\begin{figure}
\caption{ The neutron magnetic form factor normalized to the experimental
value of the neutron magnetic moment at $q^2=0$ for the momentum transfers
below $1\ GeV$. The normalization factor can be extracted from Table 1.}
\label{Figr5}
\end{figure}
\begin{figure}
\caption{ The isovector charge radius as a function of
the box size $D$ for $m_{\pi}=0$ and $m_{\pi}=139.6\ MeV$.}
\label{Figr6}
\end{figure}
\begin{figure}
\caption{ The isoscalar electric charge radius as a function of
the constituent quark mass $M$. The valence and sea parts are marked
by the dashed and dashed-dotted lines.}
\label{Figr7}
\end{figure}
\begin{figure}
\caption{ The isovector electric charge radius as a function of
the constituent quark mass $M$. The valence and sea parts are marked
by the dashed and dashed-dotted lines.}
\label{Figr8}
\end{figure}
\begin{figure}
\caption{ The electric charge radii of proton and neutron
as functions of the constituent quark mass $M$. The valence and sea
parts are marked by the dashed lines.}
\label{Figr9}
\end{figure}
\begin{figure}
\caption{ The charge density distribution of the proton (lower) and
neutron (upper) for the constituent quark mass $M =$  $420\ MeV$.}
\label{Figr10}
\end{figure}
\begin{figure}
\caption{ The magnetic moment density of proton and neutron
for the constituent quark mass $M =$  $420\ MeV$.}
\label{Figr11}
\end{figure}

\newpage
\begin{table}
\caption{ Nucleon observables calculated with the physical
pion mass.}
\vbox{\offinterlineskip
\hrule height1pt
\halign{&\vrule width1pt#&
 \strut\quad\hfill#\hfill\quad
 &\vrule#&\quad\hfill#\hfil\quad
 &\vrule#&\quad\hfill#\hfil\quad
 &\vrule#&\quad\hfill#\hfil\quad
 &\vrule#&\quad\hfill#\hfil\quad
 &\vrule#&\quad\hfill#\hfil\quad
 &\vrule#&\quad\hfill#\hfil\quad
 &\vrule width1pt#&\quad\hfil#\hfil\quad\cr
height6pt&\omit&&\multispan{11}&&\omit&\cr
& &&\multispan{11} \hfill {\bf Constituent\ Quark\ Mass} \hfill &&
&\cr
height4pt&\omit&&\multispan{11}&&\omit&\cr
&\hfill {\bf Quantity}\hfill &&\multispan3 \hfill 370\ MeV\hfill
&&\multispan3
\hfill 420   MeV\hfill
&&\multispan3
\hfill 450  MeV\hfill
&&\hfill {\bf Exper.}\hfill &\cr
height3pt&\omit&&\multispan3 && \multispan3 &&
\multispan3 &&\omit&\cr
&\omit&&\multispan{11}\hrulefill&&\omit&\cr
height3pt&\omit&&\omit&&\omit&&\omit&&\omit&&\omit&&\omit&&\omit&\cr
&\hfill    &&\hfill total \hfill &&\hfill sea \hfill &&\hfill total
\hfill &&\hfill sea \hfill && \hfill total \hfill  &&\hfill sea \hfill  &&
&\cr %
height3pt&\omit&&\omit&&\omit&&\omit&&\omit&&\omit&&\omit&&\omit&\cr
\noalign{\hrule height1pt}
height3pt&\omit&&\omit&&\omit&&\omit&&\omit&&\omit&&\omit&&\omit&\cr

& \hfill $ <r^2>_{T=0} [fm^2] $ &&
0.63  &&
0.05  &&
0.52  &&
0.07  &&
0.48  &&
0.09  &&
0.62  &\cr
height3pt&\omit&&\omit&&\omit&&\omit&&\omit&&\omit&&\omit&&\omit&\cr
\noalign{\hrule}
height3pt&\omit&&\omit&&\omit&&\omit&&\omit&&\omit&&\omit&&\omit&\cr

& \hfill $ <r^2>_{T=1} [fm^2] $ &&
1.07  &&
0.33  &&
0.89  &&
0.41  &&
0.84  &&
0.45  &&
0.86 &\cr
height3pt&\omit&&\omit&&\omit&&\omit&&\omit&&\omit&&\omit&&\omit&\cr
\noalign{\hrule}
height3pt&\omit&&\omit&&\omit&&\omit&&\omit&&\omit&&\omit&&\omit&\cr

& \hfill $ <r^2>_p [fm^2] $ &&
 0.85  &&
 0.19  &&
 0.70  &&
 0.24  &&
 0.66  &&
 0.27  &&
 0.74  &\cr
height3pt&\omit&&\omit&&\omit&&\omit&&\omit&&\omit&&\omit&&\omit&\cr
\noalign{\hrule}
height3pt&\omit&&\omit&&\omit&&\omit&&\omit&&\omit&&\omit&&\omit&\cr

& \hfill $ <r^2>_n [fm^2] $ &&
--0.22   &&
--0.14   &&
--0.18   &&
--0.17   &&
--0.18   &&
--0.18   &&
--0.12   &\cr
height3pt&\omit&&\omit&&\omit&&\omit&&\omit&&\omit&&\omit&&\omit&\cr
\noalign{\hrule}
height3pt&\omit&&\omit&&\omit&&\omit&&\omit&&\omit&&\omit&&\omit&\cr

& \hfill $\mu_{T=0}\hfill [n.m.] $ &&
0.68  &&
0.09  &&
0.62  &&
0.03  &&
0.59  &&
0.05  &&
0.88  &\cr
height3pt&\omit&&\omit&&\omit&&\omit&&\omit&&\omit&&\omit&&\omit&\cr
\noalign{\hrule}
height3pt&\omit&&\omit&&\omit&&\omit&&\omit&&\omit&&\omit&&\omit&\cr

& \hfill $ \mu_{T=1}\hfill [n.m.] $ &&
 3.56 &&
 0.77 &&
 3.44 &&
 0.97 &&
 3.16 &&
 0.80 &&
 4.71        &\cr
height3pt&\omit&&\omit&&\omit&&\omit&&\omit&&\omit&&\omit&&\omit&\cr
\noalign{\hrule}
height3pt&\omit&&\omit&&\omit&&\omit&&\omit&&\omit&&\omit&&\omit&\cr

& \hfill $ \mu_p \hfill [n.m.] $ &&
  2.12 &&
  0.43 &&
  2.03 &&
  0.50 &&
  1.86 &&
  0.43 &&
  2.79        &\cr
height3pt&\omit&&\omit&&\omit&&\omit&&\omit&&\omit&&\omit&&\omit&\cr
\noalign{\hrule}
height3pt&\omit&&\omit&&\omit&&\omit&&\omit&&\omit&&\omit&&\omit&\cr
& \hfill $ \mu_n \hfill [n.m.] $ &&
 --1.44 &&
 --0.34 &&
 --1.41 &&
 --0.47 &&
 --1.29 &&
 --0.38 &&
 --1.91    &\cr
height3pt&\omit&&\omit&&\omit&&\omit&&\omit&&\omit&&\omit&&\omit&\cr
\noalign{\hrule}
height3pt&\omit&&\omit&&\omit&&\omit&&\omit&&\omit&&\omit&&\omit&\cr
& \hfill $ |\mu_p/\mu_n| \hfill $ &&
  1.47  &&
  ---   &&
  1.44  &&
  ---   &&
  1.44  &&
  ---   &&
  1.46           &\cr
height3pt&\omit&&\omit&&\omit&&\omit&&\omit&&\omit&&\omit&&\omit&\cr
\noalign{\hrule}
height3pt&\omit&&\omit&&\omit&&\omit&&\omit&&\omit&&\omit&&\omit&\cr
& \hfill $ M_{\Delta}-M_N [MeV] $ &&
 213  &&
 ---  &&
 280  &&
 ---  &&
 314  &&
 ---  &&
 294  &\cr
height3pt&\omit&&\omit&&\omit&&\omit&&\omit&&\omit&&\omit&&\omit&\cr
\noalign{\hrule}
height3pt&\omit&&\omit&&\omit&&\omit&&\omit&&\omit&&\omit&&\omit&\cr

& \hfill $ g_A \hfill $ &&
 1.26  &&
 0.08  &&
 1.21  &&
 0.11  &&
 1.13  &&
 0.06  &&
 1.26         &\cr
height3pt&\omit&&\omit&&\omit&&\omit&&\omit&&\omit&&\omit&&\omit&\cr
\noalign{\hrule height1pt} }}
\label{Tabl1}
\end{table}

\begin{table}
\caption{ Nucleon observables calculated with the zero
pion mass.}
\vbox{\offinterlineskip
\hrule height1pt
\halign{&\vrule width1pt#&
 \strut\quad\hfil#\quad
 &\vrule#&\quad\hfil#\quad
 &\vrule#&\quad\hfil#\quad
 &\vrule#&\quad\hfil#\quad
 &\vrule#&\quad\hfil#\quad
 &\vrule#&\quad\hfil#\quad
 &\vrule#&\quad\hfil#\quad
 &\vrule#&\quad\hfil#\quad
 &\vrule width1pt#&\quad\hfil#\quad\cr
height6pt&\omit&&\multispan{11}&&\omit&\cr
& &&\multispan{11} \hfill {\bf Constituent\ Quark\ Mass} \hfill &&
&\cr
height4pt&\omit&&\multispan{11}&&\omit&\cr
&\hfill {\bf Quantity}\hfill &&\multispan3 \hfill 370\ MeV\hfill
&&\multispan3
\hfill 420   MeV\hfill
&&\multispan3
\hfill 450   MeV\hfill
&&\hfill {\bf Exper.} &\cr
height3pt&\omit&&\multispan3 && \multispan3 &&
\multispan3 &&\omit&\cr
&\omit&&\multispan{11}\hrulefill&&\omit&\cr
height3pt&\omit&&\omit&&\omit&&\omit&&\omit&&\omit&&\omit&&\omit&\cr
&\hfill    &&\hfill total &&\hfill sea &&\hfill total &&\hfill sea &&
\hfill total &&\hfill sea &&   &\cr
height3pt&\omit&&\omit&&\omit&&\omit&&\omit&&\omit&&\omit&&\omit&\cr
\noalign{\hrule height1pt}
height3pt&\omit&&\omit&&\omit&&\omit&&\omit&&\omit&&\omit&&\omit&\cr

& \hfill $ <r^2>_{T=0} \hfill [fm^2] $ &&
\hfill 0.88   &&
\hfill 0.20  &&
\hfill 0.66  &&
\hfill 0.26  &&
\hfill 0.61  &&
\hfill 0.23  &&
\hfill 0.62\hfill&\cr
height3pt&\omit&&\omit&&\omit&&\omit&&\omit&&\omit&&\omit&&\omit&\cr
\noalign{\hrule}
height3pt&\omit&&\omit&&\omit&&\omit&&\omit&&\omit&&\omit&&\omit&\cr

& \hfill $ \mu_{T=0} \hfill [n.m.] $ &&
\hfill 0.66  &&
\hfill 0.07  &&
\hfill 0.59  &&
\hfill 0.09  &&
\hfill 0.57  &&
\hfill 0.09  &&
\hfill 0.88\hfill
&\cr
height3pt&\omit&&\omit&&\omit&&\omit&&\omit&&\omit&&\omit&&\omit&\cr
\noalign{\hrule}
height3pt&\omit&&\omit&&\omit&&\omit&&\omit&&\omit&&\omit&&\omit&\cr

& \hfill $ \mu_{T=1} \hfill [n.m.] $ &&
\hfill 4.61   &&
\hfill 1.59  &&
\hfill 4.38  &&
\hfill 1.89  &&
\hfill 3.91  &&
\hfill 1.48  &&
\hfill 4.71 \hfill&\cr
height3pt&\omit&&\omit&&\omit&&\omit&&\omit&&\omit&&\omit&&\omit&\cr
\noalign{\hrule}
height3pt&\omit&&\omit&&\omit&&\omit&&\omit&&\omit&&\omit&&\omit&\cr

& \hfill $ \mu_p \hfill [n.m.] $ &&
\hfill   2.63  &&
\hfill   0.83  &&
\hfill   2.49  &&
\hfill   0.99  &&
\hfill   2.24  &&
\hfill   0.79  &&
\hfill   2.79 \hfill &\cr
height3pt&\omit&&\omit&&\omit&&\omit&&\omit&&\omit&&\omit&&\omit&\cr
\noalign{\hrule}
height3pt&\omit&&\omit&&\omit&&\omit&&\omit&&\omit&&\omit&&\omit&\cr

& \hfill $ \mu_n \hfill [n.m.] $ &&
\hfill --1.97  &&
\hfill --0.76  &&
\hfill --1.89  &&
\hfill --0.90  &&
\hfill --1.67  &&
\hfill --0.70  &&
\hfill --1.91 \hfill &\cr
height3pt&\omit&&\omit&&\omit&&\omit&&\omit&&\omit&&\omit&&\omit&\cr
\noalign{\hrule}
height3pt&\omit&&\omit&&\omit&&\omit&&\omit&&\omit&&\omit&&\omit&\cr

& \hfill $ |\mu_p/\mu_n| \hfill $ &&
\hfill 1.34 &&
\hfill ---  &&
\hfill 1.34  &&
\hfill ---  &&
\hfill 1.34  &&
\hfill ---  &&
\hfill 1.46 \hfill &\cr
height3pt&\omit&&\omit&&\omit&&\omit&&\omit&&\omit&&\omit&&\omit&\cr
\noalign{\hrule}
height3pt&\omit&&\omit&&\omit&&\omit&&\omit&&\omit&&\omit&&\omit&\cr

& \hfill $ M_{\Delta}-M_N \hfill [MeV] $ &&
\hfill 221\hfill &&
\hfill ---\hfill &&
\hfill 261\hfill &&
\hfill ---\hfill &&
\hfill 301\hfill &&
\hfill ---\hfill &&
\hfill 294\hfill &\cr
height3pt&\omit&&\omit&&\omit&&\omit&&\omit&&\omit&&\omit&&\omit&\cr
\noalign{\hrule}
\noalign{\hrule height1pt} }}
\label{Tabl2}
\end{table}

\end{document}